\documentclass[acmsmall]{acmart}
\usepackage{booktabs,multirow,listings,subcaption,wrapfig}
\usepackage{fontawesome5}   
\usepackage[most]{tcolorbox}  

\usepackage{amsmath,amssymb,amsfonts}
\usepackage{amsthm}
\usepackage{url}
\usepackage{algorithmic}
\usepackage{graphicx}
\usepackage{textcomp}
\usepackage{xcolor}
\usepackage{xspace}
\usepackage{booktabs}
\usepackage{multirow}
\usepackage{color}
\usepackage[dvipsnames]{xcolor}
\usepackage{soul}
\usepackage{listings}
\usepackage{caption}
\usepackage{subcaption}
\usepackage{wrapfig}
\usepackage{microtype}
\usepackage{array}
\usepackage{colortbl}
\usepackage{pifont}
\usepackage{fontawesome5}
\usepackage[most]{tcolorbox}
\usepackage{tabularx}
\usepackage{balance}
\usepackage[font=small]{caption}
\usepackage{hyperref}
\usepackage{lipsum}
\usepackage{balance}
\usepackage[normalem]{ulem}
\usepackage{multirow}
\usepackage{soul} 
\usepackage{color}
\usepackage{caption}
\usepackage{subcaption}
\usepackage{graphicx}
\usepackage{listings}
\usepackage{xcolor} 
\usepackage[most]{tcolorbox}
\usepackage{graphicx,wrapfig}
\usepackage{fontawesome5}

\usepackage{etoolbox}
\AtBeginEnvironment{equation}{\color{black}\small\setlength{\abovedisplayskip}{4pt}\setlength{\belowdisplayskip}{4pt}}  

\providecommand{\faThumbsDown}{\ensuremath{\blacktriangleright}}
\providecommand{\faHandPointRight}{\ensuremath{\Rightarrow}}
\providecommand{\faUnlock}{\ensuremath{\bullet}}

\newif\ifrevisionmarks
\revisionmarkstrue
\newcommand{\revise}[1]{\ifrevisionmarks{\color{black}#1}\else#1\fi}

\definecolor{anti-flashwhite}{rgb}{0.95, 0.95, 0.96}
\definecolor{lightgray}{rgb}{0.83, 0.83, 0.83}
\definecolor{dkgreen}{rgb}{0,0.6,0}
\definecolor{gray}{rgb}{0.5,0.5,0.5}
\definecolor{mauve}{rgb}{0.58,0,0.82}
\definecolor{light-gray}{gray}{0.80}
\definecolor{gray}{rgb}{0.4,0.4,0.4}
\definecolor{darkblue}{rgb}{0.0,0.0,0.6}
\definecolor{cyan}{rgb}{0.0,0.6,0.6}
\definecolor{darkred}{rgb}{128.0,0.0,0.0}

\lstdefinestyle{mystyle}{
    frame=single,
    language=Java,
    aboveskip=3mm,
    belowskip=3mm,
    showstringspaces=false,
    columns=flexible,
    basicstyle={\scriptsize \ttfamily},
    numbers=none,
    numberstyle=\tiny\color{black},
    commentstyle=\color{dkgreen},
    stringstyle=\color{mauve},
    breaklines=true,
    breakatwhitespace=true,
    tabsize=3,
    keywordstyle = {\color{blue}},
    keywordstyle = [2]{\color{blue}},
    keywordstyle = [3]{\color{mauve}},
    keywordstyle = [4]{\color{cyan}},
    otherkeywords = {changeCase, returns,exists, bool, string, then},
    morekeywords = [2]{decreases,forall, assert, print, expect}, 
    morekeywords = [3]{var, nat}, 
    morekeywords = [4]{changeCase,ChangeCase,Solution} 
}

\newtcolorbox{summarybox}[1]{
    enhanced,                         
    colback=gray!5!white,             
    colframe=black!80,                
    arc=5pt,                          
    boxrule=1pt,                      
    shadow={1.5pt}{-1.5pt}{0pt}{black!40!white}, 
    attach boxed title to top left={xshift=5mm, yshift=-3mm},
    boxed title style={
        colback=white,                
        colframe=black!80,            
        arc=3pt,                      
        boxrule=1pt,
    },
    coltitle=black,                   
    fonttitle=\bfseries,              
    title={#1},
    left=5pt, right=5pt, top=6pt, bottom=0pt
}
\newcommand{\RQthreeVerified}{14{,}934}
\newcommand{\RQthreePostCompHigh}{40.1\%}
\newcommand{\RQthreePostHarnessHigh}{35.6\%}

\newcommand{\RQthreeScoreZero}{25.7\%}
\newcommand{\RQthreeCorrectButWeak}{18.9\%}
\newcommand{\RQthreePostCorrHigh}{65.2\%}

\newcommand{\RQthreeHoudiniPreCorrZero}{90.7\%}
\newcommand{\RQthreeDaikonCorrectButWeak}{45.6\%}

\newcommand{\BestSgbHoudiniVR}{94\%}

\newcommand{\BestSgbHoudiniPreCorr}{8\%}
\newcommand{\BestSgbHoudiniPreComp}{49\%}
\newcommand{\BestSgbHoudiniPostHarness}{2\%}

\newcommand{\BestSgbSpecGenVR}{68\%}

\newcommand{\BestSgbSpecGenPostHarness}{44\%}

\newcommand{\BestFbHoudiniVR}{67\%}

\newcommand{\BestFbHoudiniPreCorr}{5\%}
\newcommand{\BestFbHoudiniPreComp}{43\%}
\newcommand{\BestFbHoudiniPostHarness}{0\%}

\newcommand{\BestFbSpecGenVR}{37\%}

\newcommand{\BestFbSpecGenPostHarness}{11\%}
\newcommand{\BestFbAutoSpecVR}{30\%}

\newcommand{\BestFbAutoSpecPostHarness}{11\%}

\newcommand{\PoolSgbSpecGenPostHarness}{44.0\%}

\newcommand{\PoolSgbAutoSpecPostHarness}{41.1\%}

\newcommand{\PoolSgbFormalBenchPostHarness}{64.4\%}
\newcommand{\PoolSgbFormalBenchNoPost}{2.0\%}

\newcommand{\PoolSgbDaikonPostHarness}{5.9\%}

\newcommand{\PoolFbSpecGenPostHarness}{28.9\%}

\newcommand{\PoolFbAutoSpecPostHarness}{25.3\%}

\newcommand{\PoolFbFormalBenchPostHarness}{42.1\%}
\newcommand{\PoolFbFormalBenchNoPost}{2.2\%}

\newcommand{\PoolFbDaikonPostHarness}{10.7\%}

\setcopyright{none} 
\renewcommand\footnotetextcopyrightpermission[1]{} 
\begin{document}
\authorsaddresses{}

\title{Spec-Harness: Assessing and Guiding LLM-Synthesized Formal Specifications Beyond Verifiability}
\author{Md Rakib Hossain Misu}
\authornote{\textbf{Corresponding author: Md Rakib Hossain Misu (mdrh@uci.edu)}}
\affiliation{
  \institution{University of California Irvine}
  \city{Irvine}
  \state{California}
  \country{USA}
}
\email{mdrh@uci.edu}
\orcid{0000-0002-7931-6782}

\author{Cristina V. Lopes}
\affiliation{
  \institution{University of California Irvine}
  \city{Irvine}
  \state{California}
  \country{USA}
}
\email{lopes@uci.edu}
\orcid{0000-0003-0551-3908}

\author{Iris Ma}
\affiliation{
  \institution{University of California Irvine}
  \city{Irvine}
  \state{California}
  \country{USA}
}
\email{huaiyaom@uci.edu}
\orcid{0009-0003-3699-7981}

\author{Yuangang Li}
\affiliation{
  \institution{University of California Irvine}
  \city{Irvine}
  \state{California}
  \country{USA}
}
\email{yuanganl@uci.edu}
\orcid{0009-0000-2259-0687}

\author{Shuvendu Lahiri}
\affiliation{
  \institution{Microsoft Research}
  \city{Redmond}
  \state{Washington}
  \country{USA}
}
\email{Shuvendu.Lahiri@microsoft.com}
\orcid{0000-0002-4446-4777}

\begin{abstract}
Formal specifications play a central role in ensuring software reliability, yet automatically synthesizing high-quality specifications remains difficult and often requires domain expertise. Recent work has applied large language models to generate specifications in the Java Modeling Language (JML), reporting high verifier pass rates. But passing a verifier only confirms that an implementation is consistent with a specification, not that the specification is meaningful: a trivial postcondition such as \texttt{ensures true} satisfies any verifier while saying nothing about the code. \revise{Mutation-based completeness scores go further, but they return a single number and cannot tell a generator which clause to fix.} In this work, we first compare classical and prompt-based JML synthesis under a unified setup, and find that \revise{22.7\% of the verified specifications have no postcondition at all, and that prompt optimization through verification feedback repairs weak prompts but does not go beyond the best hand-written one}. We then introduce Spec-Harness, a framework that measures the behavioral adequacy of a specification along four dimensions of precondition and postcondition correctness and completeness, using Hoare-triple based symbolic verification and input/output mutation \revise{on systematically built test suites}. Spec-Harness reveals that many verifier-accepted specifications are behaviorally weak, over- or under-constraining inputs and outputs in ways the verifier cannot see. Finally, we show that Spec-Harness works as a feedback signal that helps coding agents synthesize specifications with higher behavioral adequacy \revise{on held-out tests}. \revise{This holds for the general-purpose agents Codex and Claude Code and for a CodeAct-style JML synthesis agent that we built to use the verifier and Spec-Harness as feedback.}
\end{abstract}




\maketitle

\section{Introduction}

The correctness of software systems increasingly depends on the availability of precise, machine-checkable behavioral contracts. Formal specifications describe the intended behavior of programs using contracts with precise semantics, typically expressed as method preconditions and postconditions, loop invariants, or assertions at specific program locations. They form the foundation for a wide range of quality assurance tasks, including testing, model checking, and program verification. In the Java ecosystem, the Java Modeling Language (JML)~\cite{leavens2006preliminary, DBLP:conf/ifm/BoermanHJ18} provides a standard notation for writing such contracts, which tools like OpenJML~\cite{cok2011openjml} and KeY~\cite{ahrendt2014key} can then verify against the implementation. When these specifications are both correct and complete, they enable exhaustive reasoning about program behavior, detection of boundary violations, and generation of test oracles.

However, writing JML annotations in practice remains demanding, since it requires domain expertise, a deep understanding of program semantics, and significant manual effort. Rule-based approaches have been developed for automated JML synthesis~\cite{nimmer2002automatic, DBLP:conf/fm/FlanaganL01, DBLP:journals/scp/ErnstPGMPTX07}, with Houdini~\cite{DBLP:conf/fm/FlanaganL01} and Daikon~\cite{DBLP:journals/scp/ErnstPGMPTX07} being the most prominent for Java programs. Houdini uses a refutation-based approach that starts with a large set of candidate annotations drawn from predefined templates and iteratively removes those that the verifier disproves. Daikon takes a dynamic analysis approach, observing program executions to infer likely invariants over observed variable values. While both tools reduce the manual burden of writing specifications, their output is constrained by fixed templates or grammars, and this reliance on predefined patterns produces overly simple specifications~\cite{molina2022fuzzing} that fail to capture the behavior of the program.

Recent advances in LLMs have opened a promising direction for synthesizing meaningful specifications, because of their strong capabilities in code understanding and reasoning. For instance, SpecGen~\cite{DBLP:conf/icse/MaL0XB25} introduced a prompt-driven pipeline for generating JML specifications from Java method signatures and bodies, demonstrating that contemporary LLMs have sufficient syntactic and semantic understanding to produce plausible formal annotations. AutoSpec~\cite{DBLP:conf/cav/WenCSXQHLCT24} combines LLMs with static analysis by decomposing a program into components, generating specifications for each through LLM queries, and composing them into a complete annotation. It iteratively refines the result using verifier feedback until the specification is accepted. FormalBench~\cite{DBLP:conf/acl/Le-CongLM25} subsequently established a systematic benchmark for evaluating LLMs on formal specification generation, with standardized evaluation across a curated suite of Java methods and various prompts. It also proposed a mutation-based completeness score, which we discuss below.

Despite these advances, existing LLM-based approaches continue to exhibit three major drawbacks. \faThumbsDown~When an LLM is prompted to generate a specification for a given Java method, existing approaches do not check whether the model has silently altered the original code to fit the specification it produces. \faThumbsDown~More fundamentally, current approaches rely on verifier acceptance as the principal measure of quality, yet successful verification does not mean a specification is correct or complete. A trivial postcondition such as \texttt{ensures true} satisfies any verifier while saying nothing about the code's behavior, and would be useless for catching erroneous outputs during testing~\cite{DBLP:journals/pacmse/EndresFCL24, DBLP:conf/fmcad/Lahiri24, lahiri2026intent, DBLP:journals/pacmse/MisuLM024, DBLP:journals/tmlr/LoughridgeSACS025}. \faThumbsDown~Finally, telling substantive specifications apart from trivial ones still depends on manual expert evaluation, which does not scale to benchmarks with hundreds or thousands of Java methods.

\revise{FormalBench~\cite{DBLP:conf/acl/Le-CongLM25} moved past the binary verdict with its Completeness Rate (CR). It uses Major~\cite{DBLP:conf/issta/Just14} mutation tool to inject small faults into the method, verifies the specification against each faulty variant, and reports the fraction of variants that fail verification. CR is useful for comparing models, but it is a poor signal for refining a specification. It is a single number, so it cannot tell whether the precondition or the postcondition is at fault, or in which direction a clause should change. It also counts every failed verification as a detection, including timeouts and failed safety checks in the mutated code. Finally, it is defined only for specifications that already verify. In FormalBench's own results, CodeLlama-34B reaches a CR of 100\% with a success rate of 0.1\%.}

These shortcomings point to a single problem. The field measures synthesis by whether a verifier accepts the output, or at best by one aggregate score, but neither says which part of a specification is inadequate. \revise{This matters most when a generator, such as a coding agent, revises its own output: it must know which clause to change and in which direction (Section~\ref{sec:motivating}).} Progress therefore requires two things. \faHandPointRight~First, an automated way to measure how much behavior a specification captures, separately for its precondition and its postcondition, and separately for correctness and completeness. \faHandPointRight~Second, a way to feed that measurement back into synthesis, so a generator can tell which clause is too weak or too strong and revise it.

\revise{In this paper, we propose \emph{Spec-Harness}, a framework that assesses how much behavior a JML specification captures and reports it in a form a generator can act on. The idea of judging a specification by test executions is not new. Endres et al.~\cite{DBLP:journals/pacmse/EndresFCL24} and Lahiri~\cite{DBLP:conf/fmcad/Lahiri24} call a postcondition correct if it accepts the output of a valid execution and complete if it rejects altered outputs. SAFE~\cite{DBLP:conf/iclr/ChenL0GYLMYD00L25} uses this check offline to filter training data. These works judge postconditions once, after synthesis. Spec-Harness differs from them in four ways. First, it also judges the precondition, both on valid inputs and on inputs that make the method fail. Second, it reports four separate scores (PostCorr, PostComp, PreCorr and PreComp). A low score therefore points to one clause and to one direction of repair. Third, it decides every check with the verifier on a small stub without the method body, and it counts a verdict only when the verifier proves it. Fourth, it runs as a single command during synthesis, so a coding agent can use the scores as feedback while it writes the specification. The test executions come from test suites that we build systematically for each task, from its own reference implementation and sample inputs/outputs.}

We organize our study around four research questions. \revise{The first two measure synthesis by verifier acceptance, as current work does. The last two go beyond verifier acceptance with Spec-Harness.}
\begin{itemize}
  \item[\faStar] \textbf{RQ1 [Effectiveness]:} \emph{How effective are existing classical and prompt-based approaches at synthesizing verifiable JML specifications?}
  \item[\faStar] \textbf{RQ2 [Optimization]:} \emph{Can prompt optimization, leveraging structured verification feedback, improve the effectiveness of LLM-driven JML specification synthesis?}
  \item[\faStar] \textbf{RQ3 [Spec-Harness]:} \emph{To what extent do verifier-accepted specifications capture program behavior, as measured by Spec-Harness?}
  \item[\faStar] \textbf{RQ4 [Feedback]:} \emph{\revise{Can Spec-Harness be plugged into coding agents as feedback, and does it help them synthesize specifications with higher behavioral adequacy?}}
\end{itemize}

\revise{Our results show that verifier acceptance overstates what specifications capture. The best approaches verify 68.3\% of SpecGenBench and 37.6\% of FormalBench, but 22.7\% of the verified specifications have no postcondition, and 5.6\% come with code that the model changed. Optimizing prompts against the verifier lifts weak prompts, but not beyond the best hand-written one. Under Spec-Harness, only \RQthreePostHarnessHigh{} of the verified specifications both hold on most valid executions and reject most wrong outputs. With Spec-Harness feedback, all three agents write such specifications for more tasks, measured on tests that the agents never saw.}

This paper makes the following contributions:
\begin{itemize}
  \item[\faUnlock] We conduct a unified comparison of classical and prompt-based JML synthesis across six models and two benchmarks, \revise{and we test prompt optimization on top of it. We show that high verifier pass rates do not reflect actual specification quality (RQ1, RQ2).}
  \item[\faUnlock] We propose Spec-Harness, a framework that assesses the behavioral adequacy of specifications along four dimensions of precondition and postcondition correctness and completeness, using symbolic Hoare-triple checks with input/output mutation \revise{on systematically built coverage and mutation guided test suites. We use it to show that many verified specifications are weak in ways the verifier cannot see (RQ3).}
  \item[\faUnlock] We show that Spec-Harness works as a feedback signal for coding agents. \revise{Popular agents such as Codex and Claude Code can use its feedback without any change and produce specifications with higher behavioral adequacy. We also build a ReAct-style JML synthesis agent that uses the verifier and Spec-Harness as feedback. It shows that developers can build their own agent on top of Spec-Harness (RQ4).}
\end{itemize}
\section{Background \& Motivation}
\subsection{JML and Deductive Verification}
The Java Modeling Language (JML) is a behavioral specification language that allows developers to annotate Java methods with formal contracts. A JML contract consists of \texttt{requires} clauses, which define preconditions that must hold before a method executes, and \texttt{ensures} clauses, which define postconditions that must hold when the method returns. The keyword \texttt{\textbackslash result} refers to the method's return value within postconditions. For iterative code, JML provides \texttt{loop\_invariant} and \texttt{maintaining} annotations that express properties preserved across every iteration of a loop, enabling the verifier to reason about loops without unrolling them.

Once annotated, a deductive verifier such as OpenJML statically checks whether the implementation satisfies its JML specification. Internally, the verifier translates the annotated program into a set of logical proof obligations and dispatches them to an SMT (Satisfiability Modulo Theories) solver. The solver attempts to prove that each obligation holds across all possible inputs and execution paths. If all obligations are discharged, the specification is considered verified, meaning the implementation does not violate the stated contract. However, successful verification only confirms that the code is consistent with the specification. \revise{It does not show that the specification is meaningful, or how much of the method's behavior it describes.}

\setlength{\textfloatsep}{0pt}
\begin{wrapfigure}{r}{0.57\linewidth}
\vspace{-12pt}
\input{listing/motivating_example}
\vspace{-8pt}
\caption{An LLM-synthesized JML specification for the SpecGenBench task \texttt{changeCase}.}
\label{fig:motivating}
\vspace{-10pt}
\end{wrapfigure}
\setlength{\textfloatsep}{0pt}

\subsection{Motivating Example}
\label{sec:motivating}

Fig.~\ref{fig:motivating} shows an LLM-synthesized JML specification for \texttt{changeCase}, which swaps the case of a letter and returns any other character unchanged. OpenJML verifies it without errors, yet it is inadequate in two ways. The postcondition is too weak: it covers only lowercase input, and even there it bounds the result to [\texttt{'A'..'Z'}] instead of fixing it to \texttt{(char)(c - 'a' + 'A')}. For every other input, any result is allowed. The precondition is too strong: it rejects valid inputs such as \texttt{'+'} and \texttt{'\textasciitilde'}, which the method handles by returning them unchanged. Fixing the specification therefore takes two repairs in opposite directions: strengthen the \texttt{ensures} clause and weaken the \texttt{requires} clause.

\revise{Can the signals in use today point a generator to these two repairs? The verifier cannot. It accepts this specification, and it also accepts the empty specification \texttt{ensures true}, which says nothing at all. FormalBench's Completeness Rate (CR) goes further. It creates 33 mutants of the method with Major and counts the mutants for which verification fails. The specification of Fig.~\ref{fig:motivating} gets a CR of 0.36, the empty specification gets 0.09, and an exact postcondition for every branch gets 0.94. CR can therefore rank these specifications, but it does not say what to fix. Dropping the too-strong precondition repairs one of the two defects. Yet CR only rises from 0.36 to 0.39. Even this small gain does not come from the specification. One mutant fails only because OpenJML cannot prove a range check in the mutated code, and the same failure gives the empty specification its 0.09. CR also never reaches 1.0 for this method, because two mutants behave exactly like the original. A generator therefore cannot tell when it is done. Finally, if a revision makes the postcondition wrong, the specification no longer verifies, and CR is not defined at all.} Neither signal tells a generator which clause is wrong or which way to change it. To be useful as feedback, a signal has to separate the precondition from the postcondition, and correctness (the clause rejects valid behavior) from completeness (the clause admits invalid behavior). 

\section{Empirical Study: Classical vs.\ Prompt-Based Formal Specification Synthesis}
\label{sec:study}

\subsection{Baselines}
\label{sec:approaches}

We evaluate two families of approaches: classical tools built on templates or dynamic analysis, and prompt-based approaches that use LLMs (Table~\ref{tab:config}).

\par\textit{Classical approaches.} \faSave[regular]~\emph{Houdini}~\cite{DBLP:conf/fm/FlanaganL01} is a template-based annotation generator. For a Java method, it fills predefined templates with the available variables and operators to produce candidate annotations. It then calls the verifier repeatedly and removes every candidate the verifier refutes, until the remaining annotations verify.\faSave[regular]~\emph{Daikon}~\cite{DBLP:journals/scp/ErnstPGMPTX07} detects invariants dynamically. It instruments the program to trace variable values during execution, then infers likely invariants from the traces with a generate-and-check algorithm. Daikon can write its invariants as JML.

\par\textit{Prompt-based approaches.} Three LLM-based approaches for JML synthesis have been reported. All of them are prompt-based, with the verifier as their only feedback.

\faBrain~\emph{SpecGen (SG)}~\cite{DBLP:conf/icse/MaL0XB25} generates JML from method signatures and bodies with its own prompt templates. It runs a conversational pipeline in which the LLM refines its output from verifier feedback, followed by a mutation-based repair stage. We use its zero-shot (ZS), two-shot (2S) and four-shot (4S) prompts.
\par
\faBrain~\emph{AutoSpec (AS)}~\cite{DBLP:conf/cav/WenCSXQHLCT24} combines LLMs with static analysis. It decomposes the program into components, asks the LLM for a specification of each, composes them, and refines the result with verifier feedback until it is accepted. We use its ZS, 2S and 4S prompts.
\par
\faBrain~\emph{FormalBench (FB)}~\cite{DBLP:conf/acl/Le-CongLM25} is a benchmark that also provides prompting strategies for specification generation and repair with verifier feedback. We use five of them: ZS, 2S, zero-shot chain-of-thought (ZS-CoT), few-shot chain-of-thought (FS-CoT) and few-shot least-to-most (FS-LtM).

\setlength{\textfloatsep}{0pt}
\begin{wrapfigure}{r}{0.34\linewidth}
\vspace{-12pt}
\centering
\includegraphics[width=\linewidth]{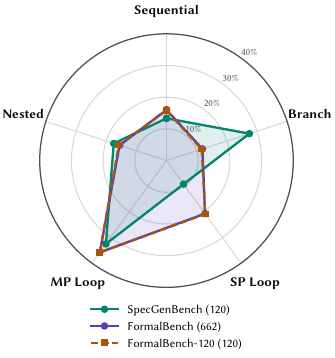}
\vspace{-16pt}
\caption{Task categories (share of tasks). FormalBench-120 is the RQ4 sample; it follows FormalBench's distribution.}
\label{fig:categories}
\vspace{-10pt}
\end{wrapfigure}
\setlength{\textfloatsep}{0pt}

\par\textit{Models.} We run every prompt-based approach with six LLMs: three proprietary models from major providers, GPT-4o~\cite{openai}, Gemini 2.0 Flash~\cite{gemini} and Claude Sonnet 4.6~\cite{claude}, and three open-weight models of similar size (32--34B parameters), Qwen2.5-Coder-32B-Instruct~\cite{qwen25coder}, DeepSeek-Coder-33B-Instruct~\cite{deepseekcoder} and CodeLlama-34B-Instruct~\cite{codellama}. The open-weight models of one size class allow a controlled comparison between them.

\begin{table}[t]
\centering
\caption{Models and prompt configurations.}
\label{tab:config}
\footnotesize
\begin{tabular}{@{}ll@{}}
\toprule
\textbf{Component} & \textbf{Configuration} \\
\midrule
Proprietary models & GPT-4o (gpt-4o-2024-08-06), Gemini 2.0 Flash, Claude Sonnet 4.6 \\
Open-weight models & Qwen2.5-Coder-32B-Instruct, DeepSeek-Coder-33B-Instruct, CodeLlama-34B-Instruct \\
\midrule
SpecGen (SG) & zero-shot (ZS), two-shot (2S), four-shot (4S) \\
AutoSpec (AS) & ZS, 2S, 4S \\
FormalBench (FB) & ZS, 2S, zero-shot CoT (ZS-CoT), few-shot CoT (FS-CoT), few-shot least-to-most (FS-LtM) \\
\midrule
Classical & Houdini, Daikon \\
\midrule
Benchmarks & SpecGenBench (120 Java methods), FormalBench (662 Java methods)\revise{, FormalBench-120 (120 tasks, RQ4)} \\
\bottomrule
\end{tabular}
\end{table}

\subsection{Dataset}
\label{sec:benchmarks}

We use two established benchmarks. SpecGenBench~\cite{DBLP:conf/icse/MaL0XB25} contains 120 Java methods from LeetCode and JML examples. FormalBench~\cite{DBLP:conf/acl/Le-CongLM25} contains 700 Java methods derived from MBJP. A small number of FormalBench methods return nothing (\texttt{void}) or use \texttt{Object} as a parameter or return type. A method without a return value cannot be evaluated through its postcondition, since there is no \texttt{\textbackslash result} to constrain, and \texttt{Object}-typed parameters lack the type information that a meaningful precondition needs. We exclude these methods, which leaves 662 FormalBench tasks. \revise{For RQ4 we also use FormalBench-120, a sample of 120 FormalBench tasks drawn per category in proportion to the full benchmark.} Both benchmarks group their tasks into five categories by control-flow structure (Fig.~\ref{fig:categories}). SpecGenBench has proportionally more branching tasks, while FormalBench leans toward loops.
\\
\textit{Normalization:} To ensure that LLMs rely solely on code reasoning rather than inferring behavior from naming conventions, we normalize all benchmark tasks by renaming class names to \texttt{Solution} and method names to \texttt{solve()}. Without this step, a class named \texttt{BinarySearch} with a method \texttt{search()} would leak semantic hints to the model, allowing it to bypass actual code comprehension. This normalization enforces a uniform evaluation setting across all approaches.

\subsection{Experiments}
\label{sec:rq1_setup}
\setlength{\textfloatsep}{0pt}
\begin{wrapfigure}[31]{R}{0.47\linewidth}
\vspace{-10pt}
\centering
\includegraphics[width=\linewidth]{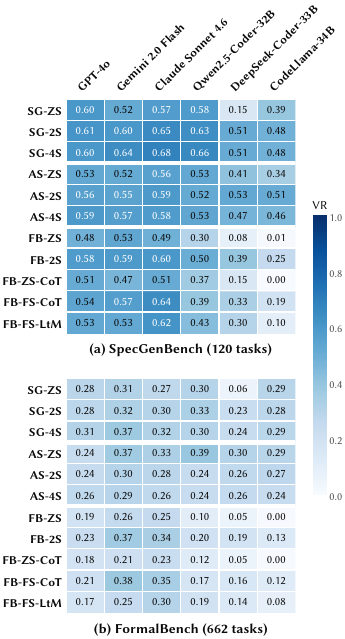}
\vspace{-14pt}
\caption{Verification Rate of the prompt-based configurations (rows: approach and prompt, e.g., SG-ZS: SpecGen, zero-shot).}
\label{fig:vr}
\vspace{-8pt}
\end{wrapfigure}
\setlength{\textfloatsep}{0pt}

We run Houdini and Daikon on both benchmarks, and the 11 prompt configurations of SpecGen, AutoSpec and FormalBench with each of the six models (Table~\ref{tab:config}), with the iteration limits of the original papers: \revise{up to 10 refinement rounds for SpecGen and AutoSpec and 3 for the FormalBench prompts, at temperature 0.7. AutoSpec was designed for C and Frama-C. We use the Java and OpenJML implementations of SpecGen and AutoSpec provided in the repository of the SpecGen authors, and the prompts of the FormalBench repository.} Daikon is the only approach that needs executions of the method: it infers invariants from the traces of the benchmark's test driver. \revise{We run it on executions that are kept separate from the ones that Spec-Harness later scores on.} This gives $2 \times 2$ classical and $11 \times 6 \times 2 = 132$ prompt-based runs, 136 in total.

Every specification is checked with the same OpenJML (21-0.21) and Java (21.0.4) setup\footnote{\texttt{openjml --esc --prover=cvc4 --nonnull-by-default --arithmetic-failure=quiet}}, with a limit of 300\,s per run. We report the Verification Rate (VR), the fraction of benchmark tasks whose specification OpenJML verifies without errors. \revise{We do not count 64 outputs that OpenJML marks as verified while reporting an internal error.} \revise{We also inspect every verified specification for two problems that VR cannot see: a missing postcondition (no \texttt{ensures} clause on the method), and code that the model changed, i.e., a renamed class, a changed signature, a modified method body or other class member, or output that is not the original method.}

\subsection{Results}
\label{sec:rq1}

\noindent\textbf{RQ1 [Effectiveness]:} \emph{How effective are existing classical and prompt-based approaches at synthesizing verifiable JML specifications?}

\par
\textit{Classical approaches.} The two classical tools show a wide gap. Houdini is the strongest baseline overall, reaching \revise{94.2\%} on SpecGenBench and \revise{66.9\%} on FormalBench (Table~\ref{tab:vr_best}), because it iteratively weakens candidate annotations against a fixed verifier. Daikon performs poorly on both (\revise{28.3\% and 24.0\%}), as its dynamic invariant inference rarely yields specifications precise enough to pass static verification.
\par
\textit{Prompt-based approaches.} On SpecGenBench, SpecGen leads at \revise{68.3\%}, followed by the FormalBench prompts at 64.2\% and AutoSpec at \revise{59.2\%}. All three best configurations use Claude Sonnet 4.6 (Fig.~\ref{fig:vr}). On FormalBench the order shifts. The FormalBench prompts are best at \revise{37.6\%}, with SpecGen at \revise{37.2\% and AutoSpec at 30.2\%. These best configurations all use Gemini 2.0 Flash.} Across both benchmarks, proprietary models beat open-weight ones \revise{(56.7\% vs.\ 37.9\% on SpecGenBench and 28.0\% vs.\ 19.8\% on FormalBench, averaged over configurations)}.
\par
\textit{Average vs.\ best configuration.} Averaged across all configurations, rates drop sharply. \revise{SpecGen falls from 68.3\% to 54.9\% on SpecGenBench and from 37.2\% to 28.3\% on FormalBench, and the FormalBench prompts fall from 64.2\% to 39.9\% and from 37.6\% to 18.7\%.} Fig.~\ref{fig:vr} makes this spread visible. The same approach swings widely across prompt types and models, so no single configuration is reliably strong everywhere. \revise{The weakest combinations nearly fail. CodeLlama-34B with the zero-shot FormalBench prompt verifies at most one SpecGenBench task.}
\par
\textit{Prompt-based vs.\ classical.} Under the best configurations, prompt-based approaches still trail Houdini (\revise{68.3\% vs.\ 94.2\% on SpecGenBench and 37.6\% vs.\ 66.9\% on FormalBench}), though they clearly beat Daikon. Under the verifier-based metric the strongest classical tool leads, but this ranking rests entirely on verifier acceptance.

\begin{table}[t]
\centering
\caption{Verification Rate of the best and average configuration of each approach, \revise{and the verified specifications (both benchmarks) without a postcondition or with changed code.}}
\label{tab:vr_best}\label{tab:vr_validity}
\footnotesize
\setlength{\tabcolsep}{4pt}
\begin{tabular}{@{}lllllrrr@{}}
\toprule
 & \multicolumn{2}{c}{\textbf{SpecGenBench}} & \multicolumn{2}{c}{\textbf{FormalBench}} & \multicolumn{3}{c}{\revise{\textbf{Verified specifications}}} \\
\cmidrule(lr){2-3}\cmidrule(lr){4-5}\cmidrule(l){6-8}
\textbf{Approach} & \textbf{Best} & \textbf{Avg.} & \textbf{Best} & \textbf{Avg.} & \revise{\textbf{All}} & \revise{\textbf{No post.}} & \revise{\textbf{Code changed}} \\
\midrule
Houdini & 113/120 (94.2\%) & -- & 443/662 (66.9\%) & -- & 556 & 93 (16.7\%) & 0 \\
Daikon & 34/120 (28.3\%) & -- & 159/662 (24.0\%) & -- & 193 & 41 (21.2\%) & 0 \\
SpecGen & 82/120 (68.3\%) & 54.9\% & 246/662 (37.2\%) & 28.3\% & 4{,}552 & 1{,}526 (33.5\%) & 516 (11.3\%) \\
AutoSpec & 71/120 (59.2\%) & 51.9\% & 200/662 (30.2\%) & 28.2\% & 4{,}476 & 1{,}599 (35.7\%) & 6 (0.1\%) \\
FormalBench & 77/120 (64.2\%) & 39.9\% & 249/662 (37.6\%) & 18.7\% & 5{,}159 & 136 (2.6\%) & 318 (6.2\%) \\
\midrule
Total & & & & & 14{,}936 & 3{,}395 (22.7\%) & 840 (5.6\%) \\
\bottomrule
\end{tabular}
\end{table}

\revise{\par\textit{What VR counts.} VR counts a specification as a success as soon as OpenJML accepts it. Table~\ref{tab:vr_validity} shows what this includes. Of the 14{,}936 verified specifications, 3{,}395 (22.7\%) have no postcondition. For 1{,}888 of them the output has no JML at all. It is the unannotated method, which the verifier trivially accepts. This is common for SpecGen and AutoSpec, where it affects about a third of the verified specifications. It is rare for the FormalBench prompts (2.6\%). 
Another 840 verified specifications (5.6\%) come with code that the model changed. In most of them (657) the model modified the method body. In total, 3{,}839 verified specifications (25.7\%) have no postcondition or change the code.}

\begin{summarybox}{Summary RQ1}
Under the Verification Rate (VR), Houdini is the strongest baseline \revise{(94.2\% on SpecGenBench and 66.9\% on FormalBench)}. The best prompt-based configurations reach \revise{68.3\% and 37.6\%}, and all prompt-based methods drop sharply from the best to the average configuration. \revise{VR also counts specifications that say nothing. Of the verified specifications, 22.7\% have no postcondition and 5.6\% come with changed code.}
\end{summarybox}
\section{Prompt Optimization for Formal Specification Synthesis}
\label{sec:gepa}

RQ1 showed that prompt-based approaches are highly sensitive to the prompt. The same approach swings widely across shot count, instruction framing and model, and the gap between the best and the average configuration is large. If the outcome depends so much on how the prompt is written, can we search for a better prompt automatically instead of writing it by hand? We ask whether prompt optimization guided by structured verifier feedback can push the verification rate beyond what fixed prompts achieve. We treat this as an open question. Optimization may close the gap, or it may reveal a ceiling that prompting alone cannot pass.

\subsection{Prompt Optimization}

Prompt optimizers search for prompt formulations that maximize a scoring function, instead of relying on hand-written prompts~\cite{DBLP:journals/corr/abs-2412-15298, agrawal2026gepa, DBLP:journals/corr/abs-2310-03714, DBLP:conf/emnlp/Opsahl-OngRPBPZ24}. Optimizers such as COPRO~\cite{DBLP:journals/corr/abs-2412-15298} and MIPRO~\cite{DBLP:conf/emnlp/Opsahl-OngRPBPZ24} work well for general NLP tasks, but they do not use feedback about \emph{why} an output failed. GEPA (Genetic-Pareto)~\cite{agrawal2026gepa} instead evolves prompts by reflection. It reads the execution traces of its attempts and the model's outputs, diagnoses the failures in natural language, proposes prompt updates, and combines lessons from the Pareto front of its attempts. This fits specification synthesis, where failures are structured. A syntax error, a failing postcondition and a type mismatch each need a different repair. GEPA uses both a numeric score and textual feedback, so it gets a direction as well as a magnitude. We therefore use GEPA as our optimizer.

\subsection{Experiments}
\label{sec:gepa_setup}

From RQ1, we select three seed prompts. They are zero-shot, few-shot chain-of-thought (FS-CoT) and few-shot least-to-most (FS-LtM). \revise{The zero-shot prompt is a plain instruction and the weakest of the three in our seed evaluation (Table~\ref{tab:gepa}). FS-CoT and FS-LtM add worked examples and step-by-step reasoning.} Using all three instead of a single strategy tests the optimizer from different starting points. For the CoT and LtM seeds, we include two SpecGenBench tasks as in-context demonstrations. We optimize on FormalBench, drawing a stratified sample by task category into a \revise{101/50/511} train/validation/test split. Stratification keeps each task category proportionally represented across the splits, which prevents the optimizer from overfitting to a narrow set of specification patterns.

\begin{wrapfigure}{r}{0.55\linewidth}
\vspace{-16pt}
\begin{minipage}{\linewidth}\small
\begin{equation}
\mathit{Score}(r) = \begin{cases}
1.0 & \text{if } r \text{ is verified}\\
0.3 & \text{if } |E_v(r)| = 1 \wedge |E_s(r)| = 0\\
0.1 & \text{if } |E_v(r)| \ge 2 \wedge |E_s(r)| = 0\\
0.0 & \text{if } |E_s(r)| > 0 \text{ or } r = \emptyset
\end{cases}
\label{eq:gepa_score}
\end{equation}
\end{minipage}
\vspace{-12pt}
\end{wrapfigure}
To give GEPA a useful signal, we replace the binary pass/fail outcome with a graduated scoring function that assigns partial credit based on failure severity. In Equation~\ref{eq:gepa_score}, $r$ is the verification result, $E_s(r)$ the set of syntax errors and $E_v(r)$ the set of verification errors. Not all failures are equal. A specification with a single postcondition violation is much closer to a correct one than a specification with a syntax error, and the optimizer should distinguish the two. GEPA also receives the classified error messages together with the score, so it gets both a gradient to climb and a direction for the repair. We implement the optimizer with DSPy's~\cite{DBLP:journals/corr/abs-2310-03714} optimizer module, using GEPA as the strategy. We use GPT-4o as both the target model, which generates the specifications, and the reflection model, which analyzes failures and proposes prompt updates. The optimization budget is set to medium. We compare each optimized prompt with its seed on the \revise{511} test tasks and on SpecGenBench (118 tasks, without the two used as few-shot examples).

\subsection{Results}
\label{sec:rq2}

\noindent\textbf{RQ2 [Optimization]:} \emph{Can prompt optimization, leveraging structured verification feedback, improve the effectiveness of LLM-driven JML specification synthesis?}

\begin{wraptable}{r}{0.5\linewidth}
\vspace{-10pt}
\centering
\caption{Verification Rate of the seed prompts and the GEPA-optimized prompts (GPT-4o), on SpecGenBench (118 tasks) and the FormalBench test split (\revise{511} tasks). \revise{For FS-LtM, GEPA kept the seed prompt.}}
\label{tab:gepa}
\footnotesize
\begin{tabular}{@{}lcccc@{}}
\toprule
 & \multicolumn{2}{c}{\textbf{SpecGenBench}} & \multicolumn{2}{c}{\textbf{FormalBench}} \\
\cmidrule(lr){2-3}\cmidrule(l){4-5}
\textbf{Seed prompt} & \textbf{Seed} & \textbf{Optimized} & \textbf{Seed} & \textbf{Optimized} \\
\midrule
Zero-shot & 39.83\% & 46.61\% & 17.81\% & 18.59\% \\
FS-CoT & 46.61\% & 44.07\% & 19.18\% & 19.77\% \\
FS-LtM & 45.76\% & 45.76\% & 19.96\% & 19.96\% \\
\bottomrule
\end{tabular}
\vspace{-8pt}
\end{wraptable}
Table~\ref{tab:gepa} compares each GEPA-optimized prompt with its seed. Optimization helps only the weakest seed: the zero-shot prompt rises from 39.8\% to 46.6\% on SpecGenBench and from 17.8\% to 18.6\% on FormalBench. \revise{For FS-CoT, the optimized prompt verifies more FormalBench tasks (19.2\% to 19.8\%) but fewer SpecGenBench tasks (46.6\% to 44.1\%). For FS-LtM, GEPA kept the seed prompt unchanged, since none of its candidates scored higher on the validation tasks. No optimized prompt verifies more than the best seed prompt on either benchmark: the best rates stay at 46.6\% on SpecGenBench (the FS-CoT seed, which the optimized zero-shot prompt only matches) and 20.0\% on FormalBench (the FS-LtM seed).} Guided by the verifier, the optimizer repairs a weak prompt, but it does not find a prompt that verifies more than the best one written by hand.

The ceiling follows from RQ1's finding rather than contradicting it. Optimization can repair a poorly written prompt, but once a prompt is good, a search guided by verifier acceptance gains little. \revise{Whether verifier-accepted specifications capture behavior at all is the question of RQ3.}

\begin{summarybox}{Summary RQ2}
Optimizing prompts against the verifier repairs the weakest prompt, but \revise{no optimized prompt verifies more than the best hand-written prompt on either benchmark}. Verifier feedback has little headroom left to give.
\end{summarybox}
\section{Spec-Harness: Evaluation Metrics for Specification Quality}
\label{sec:spec_harness}

RQ1 and RQ2 measure success only through verifier pass rates. This leaves the central question open. Do verifier-accepted specifications actually capture what the code does? A verifier only confirms that an implementation is consistent with its specification. \revise{A weak specification that constrains almost nothing therefore passes as easily as a precise one. Section~\ref{sec:motivating} also showed that neither the verifier verdict nor CR tells a generator which clause to fix.}

Recent work on verification-aware languages such as Dafny, F$\ast$ and Rust~\cite{DBLP:journals/pacmse/EndresFCL24, DBLP:conf/osdi/GoldweberYVK24,DBLP:conf/fmcad/Lahiri24, lahiri2026intent, DBLP:conf/saiv/SunSPB14, DBLP:journals/corr/abs-2507-16331,DBLP:conf/iclr/ChenL0GYLMYD00L25,swamy2026spotting} measures this gap with tests and mutation. A correct specification should hold on every valid input--output pair, and a complete one should reject incorrect outputs. The empty postcondition \texttt{ensures true} is correct but not complete, since it accepts every mutated output.

\revise{We build on this idea and design Spec-Harness around four requirements for a feedback signal. First, it must judge the precondition and the postcondition separately, because they need opposite repairs. Prior work judges only postconditions. Second, it must tell a wrong clause from a weak one. A wrong clause rejects a real behavior, and a weak clause admits a wrong behavior. Third, its verdicts must be reliable. A check that the prover cannot decide must not count against the specification. Fourth, it must be practical inside an agent loop. It should check small pieces of the specification rather than verify the whole method again for every code mutant, as mutation scores such as CR do.

Spec-Harness meets these requirements with four metrics for the correctness and completeness of the precondition and the postcondition. It never changes the method and never runs it during scoring. Instead, it replaces the method body with a stub that pins concrete inputs and outputs. It mutates only these values and asks OpenJML to prove or refute the specification on each stub. This separates it from FormalBench~\cite{DBLP:conf/acl/Le-CongLM25} and MutDafny~\cite{DBLP:journals/corr/abs-2511-15403}, which mutate the source code and verify the whole method again. To our knowledge, Spec-Harness is the first to bring this input/output check to JML, to extend it to preconditions, and to offer it to coding agents as feedback during synthesis.}

\subsection{Preliminaries}
\label{sec:prelim}
Let $P$ be a Java method with signature $m(\mathbf{x})\!:\mathbf{y}$, where $\mathbf{x}$ denotes the input parameters and $\mathbf{y}$ the return value. Let $\varphi(\mathbf{x},\mathbf{y})$ be an LLM-generated JML \emph{postcondition} and $\psi(\mathbf{x})$ an LLM-generated JML \emph{precondition} for $P$ (a missing \texttt{requires} clause means $\psi=\mathtt{true}$). Let $\mathcal{T} = \{(i_1,o_1),\ldots,(i_n,o_n)\}$ be a set of \emph{valid} input--output test pairs, i.e., pairs produced by a known-correct execution of~$P$.
\par
\revise{Two refinements follow JML semantics and the verifier configuration of RQ1. First, a postcondition constrains the result only on inputs that its precondition admits, so every check uses the \emph{guarded} postcondition $\psi(\mathbf{x})\Rightarrow\varphi(\mathbf{x},\mathbf{y})$. Without the guard, a correct postcondition would be penalized on inputs that its precondition excludes. For a specification with several cases (\texttt{also}), each case's \texttt{ensures} clauses are guarded by its own \texttt{requires}, and $\psi$ is the disjunction of the cases' preconditions. Second, OpenJML runs with \texttt{--nonnull-by-default}. Reference parameters are therefore implicitly non-null, and the \emph{effective} precondition is $\hat\psi(\mathbf{x}) = \psi(\mathbf{x}) \wedge \bigwedge_{x_j\,\in\,\mathrm{Ref}(\mathbf{x})} x_j\neq\mathtt{null}$.}
\vspace{+1mm}
\par\textit{\revise{Harness stubs.}}
For a postcondition check against test pair $(i, o) \in \mathcal{T}$, we construct a harness stub that replaces the method body with the concrete assignments $\mathbf{x} \mathrel{:=} i;\; \mathbf{y} \mathrel{:=} o$ \revise{(realized as \texttt{assume} statements that pin every parameter and a \texttt{return} of $o$)} and submits the following Hoare triple to the verifier:
\begin{equation}
  \label{eq:hoare-post}
  \models\;
  \bigl\{\,\mathtt{true}\,\bigr\}\;\;
  \mathbf{x} \mathrel{:=} i;\;\mathbf{y} \mathrel{:=} o
  \;\;
  \bigl\{\,\psi(\mathbf{x})\Rightarrow\varphi(\mathbf{x},\mathbf{y})\,\bigr\}
\end{equation}
For a precondition check against input $i$, the stub assigns $\mathbf{x} \mathrel{:=} i$ only, and the triple becomes:
\begin{equation}
  \label{eq:hoare-pre}
  \models\;
  \bigl\{\,\mathtt{true}\,\bigr\}\;\;
  \mathbf{x} \mathrel{:=} i
  \;\;
  \bigl\{\,\hat\psi(\mathbf{x})\,\bigr\}
\end{equation}
In both cases the verifier symbolically checks the Hoare triple using an SMT solver, without executing $P$. \revise{We write $\theta$ for the postcondition of the triple, that is, $\psi\Rightarrow\varphi$ or $\hat\psi$.}
\vspace{+1mm}
\revise{\par\textit{Deciding a check.} A failed proof does not show that a specification is wrong, since the prover may simply give up. Spec-Harness therefore decides every check with two proof obligations, run in parallel. The \emph{accept} obligation is the triple with postcondition $\theta$, and the \emph{refute} obligation is the same triple with $\neg\theta$. The check is accepted ($\models\theta$) or rejected ($\models\neg\theta$) when the corresponding obligation verifies, so every verdict is a proof. If neither obligation verifies within 300\,s or the tool fails, the check is \emph{undecided}. Undecided checks are left out of the denominators. When the specification's helper methods carry their own contracts, we count a check only if exactly one obligation verifies. This guards against vacuous proofs from contradictory helper contracts. We write $\mathcal{D}(\cdot)$ for the decided checks.}
\subsection{Postcondition Metrics}
\label{sec:post}

\begin{definition}[\textbf{Post-Correctness}]
\label{def:post-correctness}
A postcondition $\varphi$ is \emph{post-correct} with respect to $\mathcal{T}$ if it raises no false alarms on any known-correct execution. Formally:
{\ifrevisionmarks\color{blue}\fi
\begin{equation}
  \label{eq:post-correctness}
  \mathrm{PostCorr}(\varphi,\,\mathcal{T})
  \;=\;
  \frac{
    \bigl|\,\{\,(i,o)\in\mathcal{D}(\mathcal{T})
      \;\mid\;
      \models
      \{\mathtt{true}\}\;
      \mathbf{x}{:=}i;\,\mathbf{y}{:=}o\;
      \{\psi\Rightarrow\varphi\}
    \}\,\bigr|
  }{|\mathcal{D}(\mathcal{T})|}
\end{equation}}
\end{definition}

A score of $1.0$ indicates that $\varphi$ is consistent with every \revise{decided} valid execution in $\mathcal{T}$. \revise{On a valid input that the precondition rejects, the guarded check holds trivially. PreCorr reports this defect separately (Section~\ref{sec:pre}).} Note that a vacuous postcondition (e.g., \texttt{ensures true}) trivially achieves $\mathrm{PostCorr}=1.0$, motivating the complementary completeness metric below.
\vskip+2mm
\begin{definition}[\textbf{Post-Completeness}]
\label{def:post-completeness}
Post-completeness measures the discriminative strength of $\varphi$, that is, its ability to reject incorrect output values. \revise{For each pair $(i,o)\in\mathcal{T}$, let $\mu(o)=\{o'_1,\ldots,o'_k\}$ with $k\le 5$ be a set of \emph{output mutants} of the result type. They come from a deterministic, type-specific perturbation of $o$, and no mutant equals $o$ (Section~\ref{sec:test_data}).} Define the full mutant pool as:
{\ifrevisionmarks\color{blue}\fi
\begin{equation}
  \mathcal{T}_{1}
  \;=\;
  \bigcup_{(i,o)\,\in\,\mathcal{T}}\,
  \bigl\{\,(i,\,o')\;\mid\; o'\in\mu(o)\,\bigr\}
\end{equation}}
Let $\mathcal{T}_{2}\subseteq\mathcal{D}(\mathcal{T}_{1})$ be the subset of mutant pairs that $\varphi$ provably \emph{rejects}. \revise{A mutant counts as rejected only if the verifier proves that the precondition admits $i$ and the postcondition rules out $o'$. A failed proof of acceptance is not enough.}
{\ifrevisionmarks\color{blue}\fi
\begin{equation}
  \mathcal{T}_{2}
  \;=\;
  \bigl\{\,(i,o')\in\mathcal{D}(\mathcal{T}_{1})
    \;\mid\;
    \models
    \{\mathtt{true}\}\;
    \mathbf{x}{:=}i;\,\mathbf{y}{:=}o'\;
    \{\neg(\psi\Rightarrow\varphi)\}
  \,\bigr\}
\end{equation}}
Post-completeness is then:
{\ifrevisionmarks\color{blue}\fi
\begin{equation}
  \label{eq:post-completeness}
  \mathrm{PostComp}(\varphi,\,\mathcal{T})
  \;=\;
  \frac{|\mathcal{T}_{2}|}{|\mathcal{D}(\mathcal{T}_{1})|}
\end{equation}}
A score of $1.0$ means $\varphi$ distinguishes the correct output from all injected output faults. A vacuous postcondition scores $0.0$, as it admits every mutant unchallenged.
\end{definition}

\subsection{Precondition Metrics}
\label{sec:pre}

Precondition quality is evaluated over two disjoint input sets derived from the test suite. Let $\mathcal{T}^{+}=\{i \mid (i,o)\in\mathcal{T}\}$ denote the set of \emph{valid inputs}. $P$ returns normally on each of them. Let $\mathcal{T}^{-}$ denote a set of \emph{invalid inputs} \revise{on which $P$ fails} (e.g., \texttt{null} references, out-of-range integers, empty arrays where non-empty is required). \revise{We keep an input in $\mathcal{T}^{-}$ only if the reference implementation actually fails on it (Section~\ref{sec:test_data}).}
\vskip+2mm
\begin{definition}[\textbf{Pre-Correctness}]
\label{def:pre-correctness}
A precondition $\psi$ is \emph{pre-correct} with respect to $\mathcal{T}^{+}$ if it admits every valid input without false rejection:
{\ifrevisionmarks\color{blue}\fi
\begin{equation}
  \label{eq:pre-correctness}
  \mathrm{PreCorr}(\psi,\,\mathcal{T}^{+})
  \;=\;
  \frac{
    \bigl|\,\{\,i\in\mathcal{D}(\mathcal{T}^{+})
      \;\mid\;
      \models
      \{\mathtt{true}\}\;
      \mathbf{x}{:=}i\;
      \{\hat\psi(\mathbf{x})\}
    \}\,\bigr|
  }{|\mathcal{D}(\mathcal{T}^{+})|}
\end{equation}}
A score of $1.0$ indicates that $\psi$ imposes no spurious constraints on any valid caller. An overly restrictive precondition (e.g., one that guards against inputs that are in fact safe) is penalized here. Because the postcondition is guarded by $\psi$, such a precondition also removes those inputs from the postcondition's reach. PreCorr and PostComp therefore jointly penalize a specification that hides behavior behind a too-strong precondition.
\end{definition}
\vskip+2mm
\begin{definition}[\textbf{Pre-Completeness}]
\label{def:pre-completeness}
A precondition $\psi$ is \emph{pre-complete} with respect to $\mathcal{T}^{-}$ if it correctly guards against every known-invalid input:
{\ifrevisionmarks\color{blue}\fi
\begin{equation}
  \label{eq:pre-completeness}
  \mathrm{PreComp}(\psi,\,\mathcal{T}^{-})
  \;=\;
  \frac{
    \bigl|\,\{\,i'\in\mathcal{D}(\mathcal{T}^{-})
      \;\mid\;
      \models
      \{\mathtt{true}\}\;
      \mathbf{x}{:=}i'\;
      \{\neg\hat\psi(\mathbf{x})\}
    \}\,\bigr|
  }{|\mathcal{D}(\mathcal{T}^{-})|}
\end{equation}}
A score of $1.0$ indicates that $\psi$ rejects all invalid inputs in $\mathcal{T}^{-}$. A vacuous precondition (e.g., \texttt{requires true}) scores $0.0$ on every invalid input that is not a \texttt{null} reference, as it admits them unchallenged. \revise{If a task has no invalid inputs, PreComp is not defined for it.}
\end{definition}

\subsection{Test Data}
\label{sec:test_data}\label{sec:test_augment}

\revise{Each metric checks the specification against concrete executions of the method. PostCorr and PreCorr need valid input--output pairs. PostComp needs wrong outputs for these inputs, and PreComp needs inputs on which the method fails. The scores are only as good as these executions. If no test reaches a branch, Spec-Harness cannot expose a wrong clause about that branch.}

\begin{wraptable}{r}{0.5\linewidth}
\vspace{-10pt}
\centering
\caption{Spec-Harness metrics and the test data each one is checked on. A check counts as accepted ($\models\theta$) or rejected ($\models\neg\theta$) only when the proof obligation verifies.}
\label{tab:spec_harness_metrics}
\scriptsize
\setlength{\tabcolsep}{3pt}
\begin{tabular}{@{}llll@{}}
\toprule
\textbf{Metric} & \textbf{Checked on} & \textbf{Counts if} & \textbf{Penalizes} \\
\midrule
PostCorr & $(i,o)\in\mathcal{T}$ & $\models\,\psi\Rightarrow\varphi$ & wrong post. \\
PostComp & $(i,o')\in\mathcal{T}_1$ & $\models\,\neg(\psi\Rightarrow\varphi)$ & weak post. \\
PreCorr & $i\in\mathcal{T}^{+}$ & $\models\,\hat\psi$ & too-strong pre. \\
PreComp & $i'\in\mathcal{T}^{-}$ & $\models\,\neg\hat\psi$ & too-weak pre. \\
\bottomrule
\end{tabular}
\vspace{-8pt}
\end{wraptable}
\revise{Both benchmarks ship only a few example pairs per task (SpecGenBench: 4--34, FormalBench: 2--3). This is too few to probe a specification. We therefore build a test suite for every task automatically and without manual labelling, in three steps. First, we generate candidate inputs. They come from EvoSuite~\cite{DBLP:conf/issta/FraserZ10}, from boundary values of each parameter, and from a search that keeps an input if it kills a new Major mutant of the reference implementation. Second, we label each candidate by running the reference implementation through the benchmark's \emph{unchanged} test driver. A normal return gives a valid pair, and an exception or a timeout gives a candidate invalid input. Third, we select two \emph{disjoint} lists from each task's pool of at least 100 pairs. Both are ranked by how many new branches and killed mutants each pair adds. The \emph{evaluation} list is used for scoring, and we use its first $N{=}5$ pairs. The \emph{feedback} list is for tools that use test feedback during synthesis. Their final scores then come from tests they never saw. Invalid inputs are split between the two sides in the same way.

For PostComp, each output is perturbed into up to five values of its type, such as $o\pm1$, $o\pm2$ and $0$ for integers. On average a task has 3.4 (SpecGenBench) and 5.6 (FormalBench) invalid inputs on the evaluation side. It has none if no input makes the method fail (55 and 146 tasks). One FormalBench task (\texttt{Array3d}) has no parsable output and is excluded.

The resulting suites exercise the methods well. On SpecGenBench, the first five evaluation pairs cover 98.7\% of the branches and kill 89.1\% of the Major mutants of the method. The example pairs shipped with the benchmark cover 96.7\% and kill 81.3\%. On FormalBench, the five pairs cover 97.9\% and kill 89.3\%, against 92.3\% and 76.9\% for the shipped pairs.}

\subsection{Scoring Protocol}
\label{sec:scoring}

\revise{The four metrics score one specification on one task. To compare approaches over whole benchmarks, we need two more things. We need rules for specifications that should not get any credit, and we need a way to aggregate the scores.

\par\textit{Failed syntheses.} Two kinds of specification score $0$ on all four metrics and stay in every denominator. The first has no \texttt{ensures} clause and constrains nothing about the result. The second comes with code that differs from the benchmark program, such as a renamed class, a changed signature or method body, or code that no longer parses. The verifier then checked a different program. Adding annotations, pure helper methods or braces around a single statement is allowed.

\par\textit{Aggregation.} Approaches verify very different numbers of specifications, so mean scores over the verified specifications are hard to compare. We therefore report the fraction of synthesis attempts whose specification reaches a threshold $\tau$ on metric $M$, with $\tau$ set to 0.25, 0.5, 0.75 and 1:
\begin{equation}
  \label{eq:harness-tau}
  \mathrm{Harness}_{M}@\tau
  \;=\;
  \frac{\bigl|\{\,s \mid M(s)\ge\tau\,\}\bigr|}{|\mathit{Tasks}|\times|\mathit{Configurations}|}
\end{equation}
The denominator is the number of tasks times the number of prompt configurations. It does not depend on how many specifications verified. An attempt that did not verify counts as not reaching $\tau$. So does a metric with no decided check, which we call \emph{not evaluable}. For PreComp, the denominator counts only the tasks that have invalid inputs. Since the postcondition is what every task needs, we also report how often a specification is good on both postcondition metrics at once:
\begin{equation}
  \label{eq:post-harness}
  \mathrm{PostHarness}@\tau=\frac{\bigl|\{\,s \mid \mathrm{PostCorr}(s)\ge\tau \,\wedge\, \mathrm{PostComp}(s)\ge\tau\,\}\bigr|}{|\mathit{Tasks}|\times|\mathit{Configurations}|}.
\end{equation}
It counts the specifications that are both correct and complete on the output side. Table~\ref{tab:spec_harness_metrics} summarizes the four metrics.}


\subsection{Validating the Instrument}
\label{sec:validation}

\revise{Spec-Harness approximates adequacy with tests and mutants. Its scores therefore depend on how many test pairs it uses and on how the verifier reads a specification. We check that the scores are stable in the number of pairs and that they agree with human judgment.

\par\textit{Choice of $N$.} $N$ is the number of evaluation pairs per task. Every pair adds verifier runs. A larger $N$ thus gives more evidence but costs more, while a small $N$ may give unstable scores. We scored a stratified sample of 360 RQ1-verified specifications once with $N{=}12$ and compared smaller $N$ on the same checks. At $N{=}5$, the fraction of specifications that reach a threshold differs from $N{=}12$ by 2.1 (SpecGenBench) and 1.6 (FormalBench) points on average and by at most 5.6 points. The ranking of the approaches changes only between near-ties. Since $N{=}5$ needs about half the effort of $N{=}10$, we use it for all scoring. We read differences of up to about six points as within measurement resolution.

\par\textit{Human validation.} Two senior PhD students in computer science, each with seven years of Java experience, independently rated 100 verified specifications and 64 single checks. They did not know the approach, the model or any score. They rated the correctness and completeness of each clause and the overall adequacy. Afterwards they resolved their disagreements by discussion. Before discussion, they agreed on completeness ($\kappa = 0.93$), on adequacy ($\kappa = 0.75$) and on the single checks ($\kappa = 0.94$). Against their consensus, Spec-Harness agrees on PreCorr ($\kappa = 0.72$) and PreComp ($\kappa = 0.70$), and PostComp correlates with the completeness rating (Spearman $\rho = 0.64$). Post-Harness@0.75 matches the adequacy judgment for 67 of the 100 specifications. Most disagreements come from preconditions that reject valid inputs, and the others from limitations that we discuss in Section~\ref{sec:threats}.

\par\textit{Comparison with CR.} We also computed FormalBench's CR for the same 100 specifications. CR is defined for only 55 of them, because the other 45 do not verify under FormalBench's OpenJML options. On these 55, CR correlates with the completeness rating more strongly than PostComp ($\rho = 0.72$ vs.\ $0.45$), so it ranks specifications well. However, only 27\% of its detections come from a postcondition that rejects the mutant. The rest are timeouts (32\%) and other verification failures in the mutated code (40\%). CR thus ranks a specification without telling which clause is wrong.}

\subsection{Experiments}
\label{sec:rq3_setup}

\revise{We apply Spec-Harness to all 14{,}936 specifications of the classical and prompt-based approaches that passed the verifier in RQ1. We exclude the two specifications of \texttt{Array3d}, so 14{,}934 remain. We do not rerun the optimized prompts of RQ2. GEPA tunes a prompt for one model (GPT-4o), and no optimized prompt verifies more than the best seed prompt. All three seed prompts are already part of this set. We score each specification on the first $N{=}5$ evaluation pairs and the evaluation invalid inputs of its task. We use the verifier configuration of RQ1 with a limit of 300\,s per obligation. We report $\mathrm{Harness}_M@\tau$ per benchmark, model and approach.}

\subsection{Results}
\label{sec:rq3}

\noindent\textbf{RQ3 [Spec-Harness]:} \emph{To what extent do verifier-accepted specifications capture program behavior, as measured by Spec-Harness?}

\begin{figure}[t]
\centering
\includegraphics[width=0.88\linewidth]{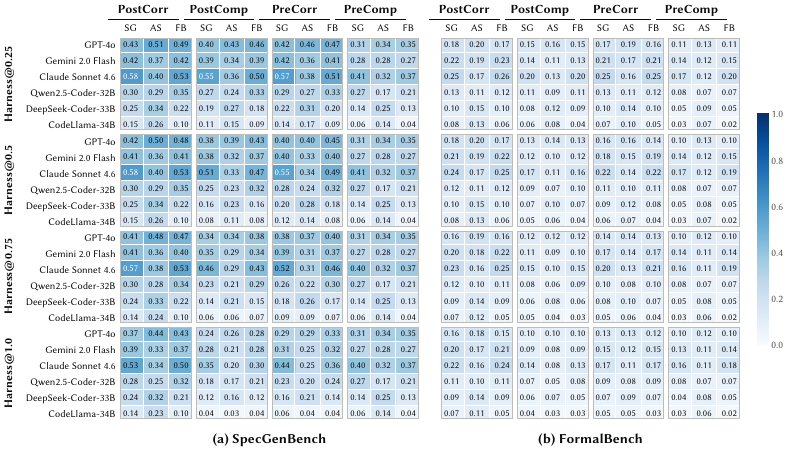}
\caption{Fraction of tasks whose specification verifies and reaches each metric at threshold $\tau$ ($\mathrm{Harness}_M@\tau$), per model and prompt-based approach (SG: SpecGen, AS: AutoSpec, FB: FormalBench prompts). \revise{PreComp: of the tasks with invalid inputs.}}
\label{fig:rq3_heat}
\end{figure}

\begin{figure}[t]
\centering
\begin{minipage}{0.88\linewidth}\centering
\includegraphics[width=\linewidth]{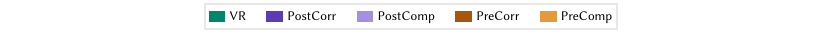}\\[1pt]
\begin{minipage}[t]{0.495\linewidth}\centering
\includegraphics[width=\linewidth]{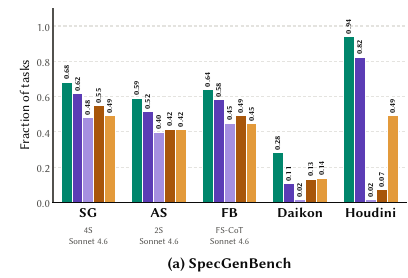}
\end{minipage}\hfill
\begin{minipage}[t]{0.495\linewidth}\centering
\includegraphics[width=\linewidth]{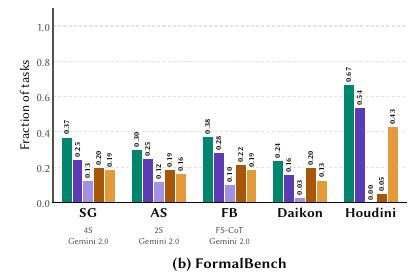}
\end{minipage}
\end{minipage}
\caption{Verification Rate next to $\mathrm{Harness}_M@0.75$ for each approach's configuration with the highest Verification Rate in RQ1 \revise{(counting only outputs that contain JML)}.}
\label{fig:rq3_vr}
\end{figure}

\revise{Fig.~\ref{fig:rq3_heat} shows the fraction of tasks whose specification verifies and reaches each metric at each threshold. It covers every model and prompt-based approach and all four thresholds. In the text, we report $\tau = 0.75$, as in the human validation (Section~\ref{sec:validation}). It requires three of every four decided checks to pass, while $\tau = 1$ fails a specification for a single check. Fig.~\ref{fig:rq3_vr} places the Verification Rate of each approach's best configuration next to its $\mathrm{Harness}_M@0.75$ scores.

\par\textit{Verification overstates adequacy.} For every approach, every Spec-Harness bar lies well below the Verification Rate (Fig.~\ref{fig:rq3_vr}). SpecGen's best configuration verifies \BestSgbSpecGenVR{} of SpecGenBench, but only \BestSgbSpecGenPostHarness{} of the tasks reach Post-Harness@0.75. On FormalBench, it drops from \BestFbSpecGenVR{} to \BestFbSpecGenPostHarness{}. Across all \RQthreeVerified{} verified specifications, \RQthreePostCorrHigh{} reach PostCorr $\ge 0.75$. Only \RQthreePostCompHigh{} reach PostComp $\ge 0.75$, and \RQthreePostHarnessHigh{} reach both. Many verified specifications hold on the valid executions but also accept wrong outputs. For example, \RQthreeCorrectButWeak{} have a perfect PostCorr and a PostComp below 0.25.

\par\textit{Specifications that say nothing.} A specification without a postcondition, or with code that the model changed, scores 0 on every metric. These are the specifications we found in RQ1 (Section~\ref{sec:rq1}), and together they make up \RQthreeScoreZero{} of the verified specifications. The extreme case is AutoSpec's zero-shot prompt with Qwen2.5-Coder-32B. It verifies 38.5\% of FormalBench, but all 255 verified outputs are the unannotated program. Fig.~\ref{fig:rq3_vr} therefore selects each approach's configuration by the Verification Rate of the outputs that contain JML. For AutoSpec on FormalBench, this is the two-shot prompt with Gemini 2.0 Flash. It verifies \BestFbAutoSpecVR{} of the tasks, and \BestFbAutoSpecPostHarness{} reach Post-Harness@0.75.

\par\textit{Classical approaches collapse.} Houdini has the highest Verification Rate (\BestSgbHoudiniVR{} and \BestFbHoudiniVR{}), but almost none of its specifications reach Post-Harness@0.75 (\BestSgbHoudiniPostHarness{} and \BestFbHoudiniPostHarness{}). Houdini removes only the candidates that the verifier refutes. A candidate precondition is never refuted, so contradictory preconditions pile up. In \RQthreeHoudiniPreCorrZero{} of its verified specifications with a postcondition, the precondition rejects every valid input. Such a specification verifies vacuously. It also rejects most invalid inputs for free, so \BestSgbHoudiniPreComp{} and \BestFbHoudiniPreComp{} of the tasks reach PreComp $\ge 0.75$. Spec-Harness exposes this through PreCorr, which only \BestSgbHoudiniPreCorr{} and \BestFbHoudiniPreCorr{} of the tasks reach. Daikon's postconditions summarize the executions it observed. For \texttt{abs}, a typical postcondition says that if the input is one of three observed values, the result is one of three observed outputs. Such postconditions hold on the evaluation pairs but constrain almost nothing. Of Daikon's verified specifications, \RQthreeDaikonCorrectButWeak{} have a perfect PostCorr and a PostComp below 0.25. Only \PoolSgbDaikonPostHarness{} and \PoolFbDaikonPostHarness{} reach Post-Harness@0.75.

\par\textit{Comparing approaches.} Pooled over models and prompts, the FormalBench prompts produce the most adequate specifications. Of their verified specifications, \PoolSgbFormalBenchPostHarness{} and \PoolFbFormalBenchPostHarness{} reach Post-Harness@0.75. SpecGen reaches \PoolSgbSpecGenPostHarness{} and \PoolFbSpecGenPostHarness{}, and AutoSpec reaches \PoolSgbAutoSpecPostHarness{} and \PoolFbAutoSpecPostHarness{}. The FormalBench prompts also rarely omit the postcondition (\PoolSgbFormalBenchNoPost{} and \PoolFbFormalBenchNoPost{}). The ranking thus differs from the ranking by Verification Rate, where Houdini leads.}

\begin{summarybox}{Summary RQ3}
\revise{Verifier acceptance overstates how much behavior a specification captures. Only \RQthreePostHarnessHigh{} of the verified specifications reach PostCorr and PostComp $\ge 0.75$, and \RQthreeScoreZero{} score 0 on every metric. Houdini has the highest Verification Rate, but it verifies mostly through preconditions that admit no valid input. Among the prompt-based approaches, the FormalBench prompts produce the most adequate specifications.}
\end{summarybox}

\section{Coding Agents with Spec-Harness Feedback}
\label{sec:agents}

\revise{RQ2 showed that optimizing against the verifier does not raise the verification rate beyond the best hand-written prompt.} RQ3 showed that verifier acceptance rewards specifications that pass, not specifications that say something. The natural fix is to change the signal. If Spec-Harness can measure whether a specification is behaviorally adequate, we can feed that measurement back into synthesis. \revise{The generator then sees when its specification is too weak or too strong, and whether the problem lies in the precondition or the postcondition. We designed Spec-Harness for this use (Section~\ref{sec:spec_harness}).} In this section we test whether Spec-Harness works as such a signal when it is offered to coding agents as a tool. \revise{We are interested in two settings. First, do popular general-purpose coding agents adopt the signal without any change and produce better specifications? Second, can a developer build their own agent around Spec-Harness with little effort? A positive answer to both would show that the benefit comes from the signal, not from a particular agent.}

\begin{figure}[t]
\centering
\includegraphics[width=0.88\linewidth]{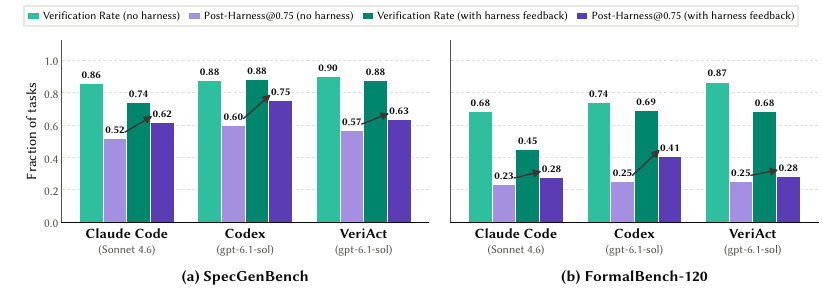}
\caption{\revise{Verification Rate and Post-Harness@0.75 of the three agents without and with Spec-Harness feedback, on the held-out evaluation suite (120 tasks per benchmark).}}
\label{fig:rq4}
\end{figure}

\subsection{Integrating Spec-Harness into Coding Agents}
\label{sec:integration}

\revise{Spec-Harness needs no change to an agent. We expose it, together with the verifier, as three command-line tools in the agent's working directory. \texttt{verify} runs OpenJML and returns its messages. \texttt{run\_specharness} prints the four scores as JSON. \texttt{submit} records the final specification. A short instruction file describes the task and the tools. The tools reject a call on code that differs from the benchmark program, as in our scoring rules (Section~\ref{sec:scoring}). We use the popular general-purpose agents Claude Code and Codex without any modification. To show that a developer can build their own agent just as easily, we also use VeriAct. VeriAct is a CodeAct-style JML synthesis agent. We built it on purpose to use the verifier results and the Spec-Harness scores as feedback.}

\subsection{Experiments}
\label{sec:rq4_setup}

\revise{We run Claude Code (2.1.284) with Claude Sonnet 4.6. We run Codex (0.159.2) and VeriAct with gpt-6.1-sol. All three use medium reasoning effort. Each agent runs in two conditions. In the \emph{No Harness} condition, it has \texttt{verify} and \texttt{submit}. In the \emph{With Harness} condition, it also has \texttt{run\_specharness}. Apart from this tool, both conditions use the same instructions and the same budget of 10 tool calls per task. The harness tool scores the specification on the task's \emph{feedback} suite (Section~\ref{sec:test_data}). The agent is asked to submit once PostCorr and PostComp reach 0.75. We use SpecGenBench and FormalBench-120 (Section~\ref{sec:benchmarks}), which gives 12 runs (3 agents $\times$ 2 conditions $\times$ 2 benchmarks). We score every final specification as in RQ3, on the evaluation suite that no agent sees. A specification counts only if it verifies with the RQ1 command, has a postcondition and leaves the code unchanged.

\par\textit{Metrics.} We report VR and Post-Harness@$\tau$ (Section~\ref{sec:scoring}). Post-Harness requires both PostCorr and PostComp. A trivial postcondition can reach a high PostCorr, and an overly strong one can reach a high PostComp, but neither reaches both. We focus on the postcondition because it is the only part of a contract that every task needs. On 46\% of the SpecGenBench and 20\% of the FormalBench-120 tasks, no input makes the method fail. For 350 methods, the two raters of Section~\ref{sec:validation} found that 70\% need no explicit precondition ($\kappa = 0.89$), and our automatic measure agrees on 99\%. As in RQ3, we use $\tau = 0.75$, which is also the agents' target. Table~\ref{tab:rq4} reports the other thresholds and metrics.}

\subsection{Results}
\label{sec:rq4}

\noindent\textbf{RQ4 [Feedback]:} \emph{Can Spec-Harness be plugged into coding agents as feedback, and does it help them synthesize specifications with higher behavioral adequacy?}

\begin{table}[t]
\centering
\caption{\revise{RQ4 results as fractions of the 120 tasks (PreComp: of the tasks with invalid inputs).}}
\label{tab:rq4}
\scriptsize
\setlength{\tabcolsep}{3.5pt}
\begin{tabular}{@{}llcccccccc@{}}
\toprule
 & & & \multicolumn{3}{c}{\textbf{Post-Harness@$\tau$}} & \multicolumn{4}{c}{\textbf{Metric $\ge 0.75$}} \\
\cmidrule(lr){4-6}\cmidrule(l){7-10}
\textbf{Agent} & \textbf{Condition} & \textbf{VR} & \textbf{0.5} & \textbf{0.75} & \textbf{1.0} & \textbf{PostCorr} & \textbf{PostComp} & \textbf{PreCorr} & \textbf{PreComp} \\
\midrule
\multicolumn{10}{@{}l}{\textit{SpecGenBench}} \\
Claude Code & No Harness & 0.86 & 0.59 & 0.52 & 0.35 & 0.81 & 0.54 & 0.68 & 0.74 \\
 & With Harness & 0.74 & 0.68 & 0.62 & 0.45 & 0.72 & 0.62 & 0.68 & 0.55 \\
Codex & No Harness & 0.88 & 0.62 & 0.60 & 0.53 & 0.74 & 0.65 & 0.72 & 0.78 \\
 & With Harness & 0.88 & 0.78 & 0.75 & 0.65 & 0.82 & 0.76 & 0.84 & 0.85 \\
VeriAct & No Harness & 0.90 & 0.62 & 0.57 & 0.47 & 0.79 & 0.62 & 0.72 & 0.83 \\
 & With Harness & 0.88 & 0.66 & 0.63 & 0.54 & 0.69 & 0.64 & 0.82 & 0.80 \\
\midrule
\multicolumn{10}{@{}l}{\textit{FormalBench-120}} \\
Claude Code & No Harness & 0.68 & 0.27 & 0.23 & 0.17 & 0.62 & 0.26 & 0.49 & 0.60 \\
 & With Harness & 0.45 & 0.35 & 0.28 & 0.23 & 0.38 & 0.33 & 0.38 & 0.36 \\
Codex & No Harness & 0.74 & 0.30 & 0.25 & 0.23 & 0.47 & 0.33 & 0.50 & 0.57 \\
 & With Harness & 0.69 & 0.49 & 0.41 & 0.35 & 0.55 & 0.43 & 0.60 & 0.64 \\
VeriAct & No Harness & 0.87 & 0.33 & 0.25 & 0.24 & 0.67 & 0.33 & 0.52 & 0.76 \\
 & With Harness & 0.68 & 0.38 & 0.28 & 0.23 & 0.41 & 0.38 & 0.57 & 0.57 \\
\bottomrule
\end{tabular}

\end{table}

\revise{\par\textit{Behavioral adequacy.} With Spec-Harness feedback, Post-Harness@0.75 rises for every agent on both benchmarks (Fig.~\ref{fig:rq4}). On SpecGenBench, it rises from 0.52 to 0.62 for Claude Code, from 0.60 to 0.75 for Codex and from 0.57 to 0.63 for VeriAct. On FormalBench-120, it rises from 0.23 to 0.28, from 0.25 to 0.41 and from 0.25 to 0.28. Compared task by task, for Codex on both benchmarks and for Claude Code on SpecGenBench, the harness turns 14 to 22 failing tasks into passing ones and loses at most three (exact McNemar test, $p \le 0.005$). For the other three runs, the gain is not significant ($p = 0.18$--$0.45$). The gain also holds at other thresholds (Table~\ref{tab:rq4}). Post-Harness rises for every agent at $\tau = 0.5$, and at $\tau = 1.0$ for all but VeriAct on FormalBench-120 (0.24 to 0.23). Codex improves on every Spec-Harness metric. This includes the precondition metrics that it was not asked to optimize. Its PreCorr $\ge 0.75$ rises from 0.72 to 0.84 and from 0.50 to 0.60.

\par\textit{Verification and cost.} Codex keeps its verification rate on SpecGenBench (0.88) and loses a little on FormalBench-120 (0.74 to 0.69). Claude Code (0.86 to 0.74 and 0.68 to 0.45) and VeriAct (0.90 to 0.88 and 0.87 to 0.68) verify fewer tasks. The specifications that do verify are much stronger. Among them, the share that reaches Post-Harness@0.75 rises for every agent. On FormalBench-120, for example, it rises from 0.34 to 0.61 for Claude Code and from 0.34 to 0.59 for Codex. A stronger postcondition is harder to verify within ten tool calls. In 37 tasks, Claude Code submitted a specification that met the harness target but did not verify. With the harness, the agents make 1.3 to 2.3 harness calls per task, and a task takes 1.4 to 2.2 times as long.

\par\textit{Held-out evaluation and integration.} In all 720 sessions, the harness returned only the four scores on the feedback suite, which shares no pair with the evaluation suite. The largest gain comes from Codex and not from VeriAct, so the benefit comes from the signal, not from a particular agent.}

\begin{summarybox}{Summary RQ4}
\revise{Spec-Harness plugs into general-purpose and custom agents alike. With its feedback, all three agents raise Post-Harness@0.75 on held-out tests by 3 to 16 points, significantly for Codex on both benchmarks and for Claude Code on SpecGenBench. Claude Code and VeriAct verify fewer specifications, but those that verify are much stronger.}
\end{summarybox}
\section{Threats to Validity}
\label{sec:threats}

\par\textit{Internal validity.} We re-ran all baselines in one environment, with the same OpenJML (21-0.21) and Java (21.0.4) versions and the same verification command throughout, \revise{including the agents' \texttt{verify} tool and the final scoring. Each agent runs once per task, so single outcomes may vary.}

\par\textit{External validity.} Our evaluation covers two Java benchmarks and six LLMs, so the results may not generalize to other languages, larger code bases or future models. \revise{Spec-Harness needs only a verifier that checks a contract on a stub with pinned values, so it can be ported to C with Frama-C.} For GEPA we used GPT-4o as both the generating and the reflecting model, which may limit the diversity of its feedback. \revise{RQ4 uses FormalBench-120 and compares each agent only with itself.}

\par\textit{Construct validity.} \revise{Spec-Harness approximates adequacy with tests and mutants. It does not cover frame conditions or exceptional postconditions. We mitigate this with coverage and mutation-guided test suites and a human validation, which also showed three limitations. OpenJML's arithmetic ignores overflow and rounding, and stubs cannot express in-place array updates. Both make a score too low. Output mutation cannot reject a postcondition listing all results of a small domain.}
\vspace{-3mm}
\section{Related Work}
\label{sec:related}

\subsection{Formal Specification Synthesis}

Before LLMs, classical tools such as Daikon~\cite{DBLP:journals/scp/ErnstPGMPTX07}, Houdini~\cite{DBLP:conf/fm/FlanaganL01} and DIG~\cite{DBLP:journals/tosem/NguyenKWF14} inferred likely invariants from runtime traces \revise{or predefined templates, but they} often produce trivial invariants (e.g., \texttt{nums != null}) and struggle with complex functional properties~\cite{DBLP:conf/icse/MaL0XB25}. LLMs shifted the field toward synthesis without exhaustive execution. Early work fine-tuned models for loop invariant inference~\cite{DBLP:conf/icml/PeiBSSY23, DBLP:conf/emnlp/ChakrabortyLFLM23}; more recent systems generate full JML specifications through mutation-guided prompting (SpecGen~\cite{DBLP:conf/icse/MaL0XB25}) or combine LLMs with static analysis (AutoSpec~\cite{DBLP:conf/cav/WenCSXQHLCT24}). Related efforts span other languages and contract types: loop invariants for C~\cite{DBLP:conf/fase/JanssenRW24}, and proof generation for Rust via agent networks~\cite{DBLP:journals/pacmpl/YangLMYCGHLLL0Z25} and self-evolving synthesis \revise{(SAFE~\cite{DBLP:conf/iclr/ChenL0GYLMYD00L25})}. Across all of these, \revise{the generator is guided and judged by verifier acceptance. SAFE goes further and filters its training specifications with test-based quality checks, but only offline, before training. Spec-Harness measures behavioral adequacy independently of the verifier and returns it to the generator while it writes the specification, as a feedback signal for coding agents.}

\subsection{\texorpdfstring{\revise{Evaluating}}{Evaluating} LLM-Synthesized Formal Specifications}

Code benchmarks such as HumanEval~\cite{DBLP:journals/corr/abs-2107-03374}, MBPP~\cite{DBLP:journals/corr/abs-2108-07732}, MBXP~\cite{DBLP:conf/iclr/AthiwaratkunGWL23} and SWE-bench~\cite{DBLP:conf/iclr/JimenezYWYPPN24} measure functional correctness by executing tests, which does not transfer to specifications, since a specification cannot be tested by running it. Code-reasoning benchmarks such as CRUXEval~\cite{DBLP:conf/icml/GuRLSS024}, CRUXEval-X~\cite{DBLP:conf/acl/XuC00LHHC025} and REval~\cite{DBLP:journals/corr/abs-2403-16437} assess input/output prediction, but still target single executions rather than \revise{complete} contracts. Efforts aimed directly at specifications include postcondition evaluation~\cite{DBLP:journals/corr/abs-2407-14118}, \revise{the effect of fine-tuning} on formal-methods benchmarks~\cite{DBLP:conf/acl/Cao0LMLH000QCT25}, SpecEval~\cite{DBLP:journals/corr/abs-2409-12866} (preconditions, postconditions and loop invariants with counterfactual analysis), and FormalBench~\cite{DBLP:conf/acl/Le-CongLM25} (Java methods spanning diverse execution behaviors). \revise{Most of them score a specification as correct or incorrect against a ground-truth or verifier oracle. FormalBench adds a mutation-based completeness rate, but it is one score per specification and is defined only once the specification verifies (Section~\ref{sec:motivating}). Closest to our work, Endres et al.~\cite{DBLP:journals/pacmse/EndresFCL24} and Lahiri~\cite{DBLP:conf/fmcad/Lahiri24} judge LLM-generated postconditions by whether they hold on valid tests and reject buggy behavior. Spec-Harness brings this idea to JML and its verifier, adds the precondition, and separates correctness from completeness on both sides, so that each deficiency is visible and points to a repair.}
\vspace{-3mm}
\section{Conclusion}
\label{sec:conclusion}
We studied the gap between verifiable and behaviorally adequate formal specifications. \revise{Our study of classical and prompt-based JML synthesis showed that verifier pass rates also count specifications that say nothing (RQ1). Optimizing prompts against the verifier repaired weak prompts but did not go beyond the best hand-written one (RQ2). To look past verifier acceptance, we introduced Spec-Harness. It checks the precondition and the postcondition separately, for both correctness and completeness, and decides every check with a proof. Only \RQthreePostHarnessHigh{} of the verified specifications reach PostCorr and PostComp $\ge 0.75$. Houdini has the highest Verification Rate, but it verifies mostly through preconditions that admit no valid input (RQ3). With Spec-Harness as feedback, Codex, Claude Code and our ReAct-style agent produce specifications that capture more behavior on held-out tests (RQ4). Our results suggest that work on specification synthesis should report behavioral adequacy next to the verification rate, and that verifier feedback alone is not enough to guide synthesis. In future work, we plan to port Spec-Harness to other language popular verifiers such as Frama-C, Verus/Rust to compare its effectiveness.}

\section*{Data Availability}


All artifacts associated with this study are available here in the \href{https://github.com/Mondego/vACT}{\faGithub~\texttt{vAct}} GitHub Repository. It includes implementation and execution scripts for all baseline approaches (Daikon, Houdini, SpecGen, AutoSpec, and FormalBench), normalized benchmark datasets (SpecGenBench and FormalBench) with extended test suites, our implementations of the GEPA optimizer, Spec-Harness, and VeriAct.

\balance
\bibliographystyle{ACM-Reference-Format}
\bibliography{references}

\end{document}